\documentclass[]{raa}
\usepackage{CJK}

\usepackage{graphicx,times}
\usepackage{natbib}
\usepackage{amssymb,amsmath}
\bibpunct{(}{)}{;}{a}{}{,}

\usepackage[pagebackref=true]{hyperref}

\begin{document}
\begin{CJK*}{UTF8}{gbsn}

\title{Can near-to-mid infrared spectral energy distribution quantitatively trace protoplanetary disk evolution?}

 \volnopage{ {\bf 20XX} Vol.\ {\bf X} No. {\bf XX}, 000--000}
   \setcounter{page}{1}

\author{Mingchao Liu (刘明超)\inst{1,2,3}, Jinhua He (何金华)\inst{1,2,4}, Zhen Guo\inst{5,6}, Jixing Ge (葛继兴)\inst{7,8}, Yuping Tang (唐雨平)\inst{9}}

\institute{Yunnan Observatories, Chinese Academy of Sciences, 396 Yangfangwang, Guandu District, Kunming, 650216, Yunnan, P. R. China; {\it lmch@ynao.ac.cn; jinhuahe@ynao.ac.cn}\\
\and
Chinese Academy of Sciences South America Center for Astronomy, National Astronomical Observatories, Chinese Academy of Sciences, Beijing, 100101, P. R. China\\
\and
University of Chinese Academy of Sciences, Yuquan Road 19, Shijingshan Block, Beijing, 100049, P. R. China\\
\and
Departamento de Astronom\'{i}a, Universidad de Chile, Las Condes, 7591245 Santiago, Chile\\
\and
Instituto de F{\'i}sica y Astronom{\'i}a, Universidad de Valpara{\'i}so, ave. Gran Breta{\~n}a, 1111, Casilla 5030, Valpara{\'i}so, Chile\\
\and
Centre for Astrophysics Research, University of Hertfordshire, Hatfield AL10 9AB, UK\\
\and
Xinjiang Astronomical Observatory, Chinese Academy of Sciences, Urumqi 830011, China\\
\and
Xinjiang Key Laboratory of Radio Astrophysics, 150 Science1-Street, Urumqi 830011, China\\
\and
Shanghai Key Lab for Astrophysics, Shanghai Normal University, 100 Guilin Road, Shanghai 200234, Peopleʼs Republic of China
\vs \no
   {\small Received 20XX Month Day; accepted 20XX Month Day}
}



\abstract{
Infrared (IR) spectral energy distribution (SED) is the major tracer of protoplanetary disks.
It was recently proposed to use the near-to-mid IR (or K-24) SED slope $\alpha$ defined between 2-24$\mu$m as a potential quantitative tracer of disk age.
We critically examine the viability of this idea and confront it with additional statistics of IR luminosities and SED shapes.
We point out that, because the statistical properties of most of the complicated physical factors involved in disk evolution are still poorly understood in a quantitative sense, the only viable way is to assume them to be random so that an idealized `average disk' can be defined, which allows the $\alpha$ histogram to trace its age. 
We confirm that the statistics of the zeroth order (luminosity), first order (slope $\alpha$) and second order characteristics (concavity) of the observed K-24 SEDs indeed carry useful information upon the evolutionary processes of the `average disk'.
We also stress that intrinsic diversities in K-24 SED shapes and luminosities are always large at the level of individual stars so that the application of the evolutionary path of the `average disk' to individual stars must be done with care.
The data of most curves in plots are provided on GitHub\footnote{Disk-age package \url{https://github.com/starage/disk-age/}}.
}

   \authorrunning{Mingchao Liu et al. }            
   \titlerunning{Can K-24 SED quantify disk evolution?}  
   \maketitle



\section{Introduction} 
\label{sec:introduction}

The formation of low and intermediate mass stars has been intensively studied and a standard picture of star formation (SF) has been there since a long time ago \citep[e.g.,][]{Shu1987,Larson2003,McKee2007,Evans2009}. 
However, despite of tremendous efforts on observational studies \citep[e.g.,][among many others]{Myers1987,Palla2000,Haisch2001,Evans2009,Mamajek2009}, our quantitative knowledge of the star formation process is still limited \citep[e.g.,][]{Krumholz2011}.
More quantitative observational constraints on the earliest disk evolutionary stages are highly desired to constrain disk evolution models, either at individual star level \citep[e.g.,][among many other models]{Zhu2010,Kimura2016,Kunitomo2020,Kunitomo2021}{}{} or at population level \citep[e.g.,][]{Emsenhuber2023}{}{}, and to test models of planet formation \citep[e.g.,][]{Mordasini2009,Pignatale2018,Burn2021,Raymond2022}.

Although there have been many works to quantify disk evolution through the observation of their host stars \citep[see e.g., the review of][]{Manara2023}{}{}, the most straightforward method should be using the IR emission of the disks themselves. 
The first way is to utilise the IR colors in the color-color diagram or color-magnitude diagram which is similar to the Hertzsprung-Russell diagram \citep[HRD; e.g.,][]{Palla2000} for stars.
However, the colors were usually used to identify YSOs, but not to quantify their evolution \citep[e.g.,][among many others]{Luhman2008,Rebull2010}{}{}.
The color-magnitude diagram is also known to suffer from large uncertainty during the earliest millions of years (Myrs) when the primordial disks exist \citep{Soderblom2014}.

The second way is to use the characteristics of the IR SED.
The most frequently used parameter is the SED slope defined in logarithmic scale \citep[known as $\alpha$;][]{Lada1987,Adams1987,Greene1994}.
Since the launch of the Spitzer Space Telescope \citep{Werner2004}, both the IR colors and SED slope have been intensively used to explore the properties of nearby populations of YSOs \citep[e.g.,][etc.]{Luhman2008,Rebull2010,Luhman2010,Kirk2009,Megeath2012,AlvesdeOliveira2012,Hsieh2013,Broekhoven2014,Dunham2015,Azimlu2015}.
However, most of the existing works follow the classical framework of disk evolution described in the IR SED classification scheme of {\rm\textsc{0, i, f, ii} and \textsc{iii}} \citep[][]{Lada1987,Adams1987,Andre1993,Greene1994}{}{}.
A typical example is to refine the lifetime estimation of each SED classes using improved star counts \citep[e.g.,][]{Wilking1989,Evans2009}{}{}.

In a recent work, \citet{Liu2023} proposed an innovative idea that the near-to-mid IR SED (or K-24 SED in short) slope $\alpha$ defined between 2-24$\mu$m can be directly used to quantify the disk evolutionary age in a novel manner not confined by the classical SED classification scheme. 
Their key point is the striking similarity of the $\alpha$ histograms among the 13 largest nearby protoclusters in Gould's Belt (GB). 
This also triggered them to propose that the large sample of disk SEDs carry the information of disk evolution timescales and one only has the opportunity to examine the star formation histories of individual protoclusters after the common disk evolution timescales have been taken into account.
Indeed, only few works in literature paid attention to the peaks of the $\alpha$ histograms, speculated that the peaks might reflect episodic SF histories \citep[e.g.,][]{Hsieh2013}{}{}.
However, the similarity of the $\alpha$ histogram among all the GB protoclusters cannot be easily reconciled by episodic SF events, because there is no known mechanism to synchronize these parsec-scale events across so large a distance up to a kilo-parsec, which is spanned by the entire Gould's Belt.

With the belief that their average $\alpha$ histogram does trace disk evolution, \citet{Liu2023} went further to divide the intrinsic disk evolution processes into five stages, A, B, C, D and E, according to the characteristics of the histogram. 
They also discussed how their staging scheme can be better matched to known physical processes of disk evolution than the traditional SED classes do.

From the theoretical point of view, it is reasonable to imagine an evolutionary sequence of disk structure for a single YSO and it is fair to allow the disk to have its own evolutionary age. 
As the major tracer of disks, the IR SED or more specifically its SED slope $\alpha$, does have the potential to serve as an independent age tracer for disks, just as the HRD does for the central stars.
However, as we will discuss below, the disk evolution involves complicated physical processes and randomness. 
It is yet uncertain how similar the SED evolution paths would be among YSOs formed in different initial conditions and different local environments.
The proposal of \citet{Liu2023} is equivalent to assuming that these complication and randomness can be effectively suppressed by the use of large star samples so that their averaged $\alpha$ histogram mainly reflects the intrinsic disk evolution.
But a scrutiny on this point is necessary before we know to what extent this disk-age tracer can be trusted.
For this purpose, we first give a brief review of the complicated disk evolution processes below. 

\subsection{The complicated relationship between protoplanetary disk evolution and K-24 SED}
\label{subsec:overview}

The relationship between the evolution of protoplanetary disks around low mass stars and their K-24 SEDs is expected to be complicated. 
In the earliest star-formation stage, the accretion disk is deeply embedded in a gas envelope which is the inner part of its natal molecular cloud dense core. 
The SED of such an object is very red and usually too faint to be observable in IR due to high extinction. 
The first opportunity to observe the youngest disk in IR is when the stellar accretion has generated high speed bipolar jets to carve out a pair of bipolar cavities in the envelope.
The strength and shape of the observed K-24 SED in this stage sensitively depend on the inclination angle of our sight line with respect to the bipolar cavities and the detailed properties of the latter, and both thermal emission and scattered stellar light can contribute \citep[e.g.,][]{Kenyon1993,Whitney1993,Whitney2003}.
After then, due to the action of photoevaporation and shock pressure of the jets, the bipolar cavities may expand quickly and the envelope around the disk will be cleared \citep[see some evidences in][]{Jorgensen2009,Furlan2016}. 
The K-24 SED in this stage still sensitively depends on the inclination angle and evolves with time significantly.

Later on, the dust and gas in the disk may settle down into a layered structure \citep[e.g.,][]{Dullemond2004}. 
Cold dust grains in the middle plane of the disk may aggregate into bigger ones to seed the formation of protoplanets \citep[e.g.,][]{DAlessio2001,DAlessio2006}, while the disk gas may continue escaping from the surface layers through photoevaporation by FUV, EUV and X-ray light from the central accreting star \citep{Clarke2001,Alexander2006,Rosotti2013} and maybe also due to magnetohydrodynamic disk winds \citep[e.g.,][]{Bai2016}.
The inner disk edge may be puffed up by UV light heating to form an inner wall to protect part of the disk against photoevaporation through shadowing \citep{Dullemond2004}.
The K-24 SED of such a disk is determined by the thermal structure of the disk that is regulated by various disk parameters such as the height, shape and thickness of the inner wall and the flaring angle of the disk. 
The inclination angle may still play some role.
Formation of planets, particularly giant gaseous planets, in this stage may induce dynamical features such as spirals, multiple gaps of diverse widths or even a large inner hole in the disk \citep[a transitional disk; e.g.,][]{Alexander2006,Varniere2006,Gorti2009,Dodson-Robinson2011}, which may have some impact to the K-24 SED. 
If a multiple star system is formed, the powerful perturbation of the secondary star may truncate a significant part of the disk or even disperse the entire disk earlier, resulting in a strong modification to the evolution path of K-24 SED \citep{Kraus2012}.
Finally, the primordial disk will be dispersed entirely by photoevaporation so that the contribution of disk emission to K-24 SED diminishes. 

In addition to the complexity, the disk evolution process also involves huge randomness.
For example, the initial envelope and disk masses, the initial inner disk radius (depending on stellar mass), the extinction of parent molecular cloud, the duration of accretion phase, and the number, masses and radii of planets formed in the disk \citep[see e.g.,][]{Robitaille2006} are all random to some extent.
The disk accretion to central star is also intermittent \citep{Kenyon1990}, resulting in temporal variation of intrinsic K-24 SED shape and also additional uncertainty in observed SED shape when the different photometric bands would not be observed simultaneously.
Particularly, the random inclination angle of the disk are believed to be able to alter observed K-24 SEDs significantly.  \citep{Whitney2003, Crapsi2008}.

In this paper, we will first discuss how the above complexity would impact the capability of the $\alpha$ histogram in quantifying disk evolution in Sect.~\ref{section:Impact of physical factors}.
Then, we report our scrutiny of the $\alpha$ histogram of \citet{Liu2023} in Sect.~\ref{sec:alpha-histo}.
To strengthen our query into the reliability of the SED slope (a first order characteristics of the SED) as a disk age tracer, we will further employ the statistics on the second order SED characteristics (the IR SED shape) and zeroth order SED characteristics (IR luminosities in the Spitzer IRAC and MIPS\,$24\mu$m bands) of the same YSO samples in Sect.~\ref{sec:Diversity} and \ref{sec:luminosity-trends}, respectively.
Particularly, we will be able to discern whether the most important factor, the contamination from edge-on opaque disks, could prevent the IR SED from being an effective age indicator.
Finally, after a dedicated discussion of the uncertainties in the $\alpha\sim-1$ histogram peak in Sect.~\ref{sec:Remaining SED randomness}, we draw our conclusions in Sect.~\ref{sec:summary}.

\section{Impact of uncertainties to the average SED-slope histogram
} 
\label{section:Impact of physical factors}

For many complicated processes and random factors discussed in Sect.~\ref{subsec:overview}, we still do not have enough knowledge of them at the level of individual YSOs so that their impact to K-24 SED have to be considered as random processes at population level.
This is the case for the distribution of initial conditions (e.g., initial envelope and disk masses, inner disk radius), total length and intermittency of accretion phase, diversity of disk structures such as inner wall, disk layers and flaring factor, the dust grain settling and growth, planet formation and planet-disk interaction, extinction and observational incompleteness.
When a large population of YSOs are considered, as we are discussing here, we can expect these random factors to be greatly suppressed through averaging. 
However, concerning the average $\alpha$ histogram of \citet[][their Fig.~1; also see an improved version in our Fig.~\ref{fig: alpha-age} in Appendix]{Liu2023}, the sample sizes of the earliest stage A, the earlier part of Stage B and the latest stage E are not large enough so that the advantage of averaging does not work.
Fortunately, the earliest stages are brief so that they only have limited impact to age estimation in later stages.
Furthermore, we should also remain alert in future on possible distortion of the $\alpha$ histogram by any yet unrecognized non-negligible parameter correlations.
Below, we briefly discuss several better understood factors among them.

The intermittent disk accretion has been studied since a long time ago \citep[e.g.,][]{Kenyon1990,Hartmann1996,Megeath2012,Hartmann2016} and has become a hot topic \citep[e.g.,][]{Vorobyov2015,Caratti2017,Contreras2017,Guo2020,Guo2024a,Guo2024b,Park2021}. 
But the temporal behavior of such bursts is still not well predictable \citep{Fischer2023}.
There have been clues indicating that the intermittency could be related to random gas in-fall onto the disk \citep{Kuffmeier2018}. 
Therefore, currently, the best choice is to consider the accretion bursts as a random process.

Another important complication, the gaps, central hole or other dynamical effects created in disks by protoplanets or stellar companions, can also be treated as random. 
Despite of a long history of intensive modeling and observations of planet formation and its interaction with the disk \citep[e.g.,][among many others]{Pollack1996,Alibert2005,Rosotti2013,Baruteau2014,Dong2015a,Dong2015b,Andrews2018,Long2018,Emsenhuber2021}{}{}, the number of planets, their masses and orbit radii, or similar parameters of stellar companions, are still hard to predict for a real population of protoplanetary disks. 

Differential extinction can modify observed K-24 SEDs. 
Generally, only foreground extinction of interstellar medium can be easily corrected in observations. 
The remaining, more localized extinction can be classified into two categories: 1) the one due to local cloud and circumstellar envelope and 2) the one due to disk.
The former is known to be larger in denser clouds where the younger YSOs of Stages A and B reside \citep[e.g.,][among many others]{Getman2014,Azimlu2015}.
However, in any evolutionary stage of disk or in any given cloud region, its value for different YSOs is more or less random.
Beside the randomness, the remaining deterministic effect of K-24 SED reddening indeed has only limited impact to the $\alpha$ histogram due to the paucity of objects in these early stages.

The second category of extinction (by disk) sensitively depends on the geometrical and optical thickness of the disk and some secondary parameters such as flaring and self-shadowing by inner wall. 
Its impact is expected to be large among Class 0, \textsc{i} and \textsc{f} YSOs in the earliest Stages A and B when the disk is geometrically thick.
Even for later Class \textsc{ii} objects in Stages C and D, \citet{Crapsi2008} had suggested that the impact of disk inclination angle would be as high as $30\%$. 
However, we would like to figure out an intriguing difference between the inclination angle effects and the contamination from edge-on opaque disks (opaque disks that totally block out the inner hot disk region): while the former is likely always important, the latter is not necessarily so.
For example, if the disks are always very opaque during most of their lifetime, as demonstrated by the model grid of \citet{Robitaille2006}, edge-on disks may become so faint at near IR (NIR), due to strong extinction, that they almost never appear in our Spitzer/IRAC selected sample.
In this case, the edge-on opaque disks do not contribute to modifying observed $\alpha$ values, while the inclination angle effect is still important in determining the fraction of observable disks (depending on the cavity opening angle) and in reddening the observed SEDs to a smaller extent (depending on structural details of the disk surface) in all evolutionary stages.
As we will further explore in more detail in the next sections, the real impact of edge-on disk contamination is indeed not dominant in the GB samples.

Observational incompleteness of YSOs at population level can have strong impact to $\alpha$ histograms too. 
The main sources of incompleteness are the confusion with background/foreground main sequence stars \citep[e.g.,][]{Harvey2007}{}{} and high extinction. 
The former mainly affects the Class\,\textsc{iii} objects in Stage E due to the similarity of their SEDs to that of stellar photosphere, while the latter leads to the missing of YSOs with lower luminosity.
This fact reinforces the vulnerability of the average $\alpha$ histogram in the earliest and latest stages A, early B, and E.
Perhaps some on-going projects such as SESNA \citep{Pokhrel2020} could improve it to some extent with a better way of source extraction in the Spitzer maps.

In a summary, while the randomness in the disk evolution processes could be suppressed by using large samples of YSOs, the relationships between disk parameters are still poorly known and 
there is some risk to treat all the disk parameters as independent random factors. Thus, further testing of the reliability of the SED slope as a disk age tracer will be beneficial.

\section{Scrutinize the K-24 SED-slope histogram}
\label{sec:alpha-histo}

As the first step of our scrutiny, we closely examined the methodology of \citet{Liu2023}.
Although their judgement of cluster membership, new extinction correction, handling of Spitzer MIPS $24\,\mu$m saturation cases and the definition of K-24 SED slope $\alpha$ are acceptable, we identify two important improvements. 
The details of the improvements, an updated version of the average $\alpha$ histogram and a simple discussion of its potential to trace disk evolutionary age are given in Appendix~\ref{subsec:alpha_histograms}.
The main conclusion is that it is mainly the Stages B, C and D of \citet{Liu2023} that potentially have the power to trace disk ages. 

We also compare the observed $\alpha$ histogram to that of the disk model grid of \citet{Robitaille2006} in Appendix~\ref{subsection:SED slope and age} in order to verify whether the former contains useful information of disk evolution to constrain disk models.
Despite of a number of important limitations of the model grid \citep[Sect.~2.2.4 of][]{Robitaille2006,Robitaille2008}, it has integrated the disk physics known at that time.
A simple experiment is also performed by constraining the central star mass of the models to $M^*<1$\,M$_\odot$ in order to verify the potential of the observed $\alpha$ histogram in constraining disk models.
It is found that the observed and the modeled $\alpha$ histogram are very different. 
The tested central star mass is a sensitive parameter that can be hopefully constrained using the observed $\alpha$ histogram in future works. Therefore, we conclude that
the observed $\alpha$ histogram can provide useful constraints to disk model grids.

\section{Test the evolution of K-24 SED shape and its diversity}
\label{sec:Diversity}

Real SEDs usually take curvy shapes, which is jointly determined by the multiple factors discussed above. 
If the SED slope $\alpha$ really can trace disk evolution, we should expect that the change of the average SED shape across the bins of the $\alpha$ histogram mainly reflects the evolution of disk structure from a spherical envelope to bipolar cavities to a geometrically thick disk and finally to a geometrically thin disk, while the diversity of the SED shapes measures the randomness of the physical factors.
If the contamination from edge-on opaque disks would have played a major role during any of the Stages B, C or D, we should expect more concave SED shapes and an enhancement of the SED-shape diversity at larger $\alpha$ value, due to higher extinction at shorter wavelengths and due to the contribution of scattered light in optical and NIR, according to the simulations of \citet{Whitney2003,Robitaille2006} etc.
We can verify these trends among the same sample of GB YSOs.

\begin{figure}[ht!]
\centering
\includegraphics[width=6.0cm, angle=0]{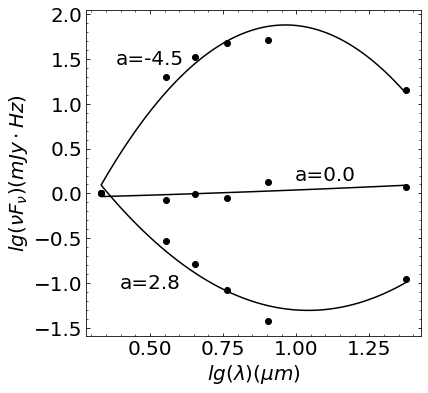}
\caption{Examples of quadratic polynomial fitting (full line) of observed SEDs (dots) to illustrate how the second order coefficient $a$ quantifies the SED concavity.
The observed SEDs have been shifted to the same value of zero in K$_{\rm s}$ band (left most wavelength) for clarity.}
\label{fig:fit_concavity}
\end{figure}

To character the SED shape, we extend the linear fit to the SEDs when defining the slope $\alpha$ to a quadratic one
\begin{equation}
    \label{eq: concavity}
    lg(\lambda F_{\lambda})=alg^{2}(\lambda)+blg(\lambda)+c
\end{equation}
so that we can use the second order coefficient $a$ to characterize the shape (or concavity) of an SED.
An SED is overall convex when $a<0$, lacking second order concavity when $a=0$, overall concave when $a>0$.
An example of each case is shown in Fig.~\ref{fig:fit_concavity}.
Note that this concavity does not reflect the Silicates feature around $9.8\mu$m.
The coefficient $b$ is similar to $\alpha$, but we still use the latter to represent the SED slope, because it is more consistent with the tradition in literature.

The SED shape reflects the joint effects of the wax and wane of the major emission components (e.g., the central star, inner and outer parts of the disk, bipolar cavities and envelope) and extinction.
Previous works used SED slopes or colors over diverse wavelength ranges such as $3.6-8.0\mu$m and $5.8-24\mu$m \citep[e.g.,][]{Fang2009,Fang2013}{}{} or a turnoff wavelength $\lambda_{\rm turnoff}$ of the SED \citep{Cieza2007} to characterize the SED shape. 
The second order coefficient $a$ in our Eq.~\ref{eq: concavity} can quantify such information in the simplest manner enabling statistical study.
\begin{figure*}[ht!]
\centering
\includegraphics[width=12.0cm, angle=0]{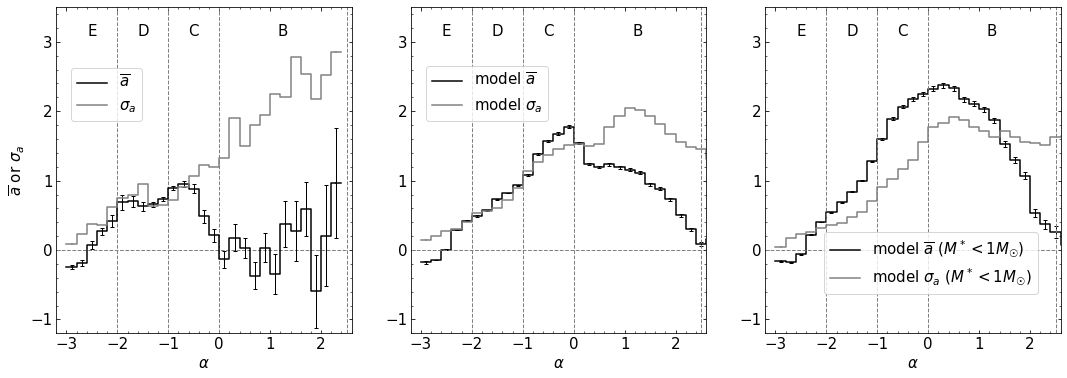}
\caption{The average ($\overline{a}$, black, the error bar is the uncertainty) and sample standard deviation ($\sigma_{a}$, gray) of the SED-shape parameter $a$, as defined in Eq.~\ref{eq: concavity}, for each $\alpha$ bin of width $\Delta\alpha=0.2$ \citep[left for all the 5194 observed samples in Gould's Belt; middle and right for the entire model grids of][]{Robitaille2006}{}{} and a sub-grid with the central star mass confined to $M^*<1$\,M$_\odot$, respectively, with the model SEDs averaged over the 50 model apertures). 
Only the bins with observed star number $\ge10$ are shown. 
The long vertical gray dashed lines mark the border lines of the disk evolution Stages B, C, D and E of \citet{Liu2023}. The horizontal gray dashed line at $Y=0$ is for comparison.
The data of the observed and full grid model curves are given on the GitHub mentioned in Abstract.
\label{fig:std and average of a}}
\end{figure*}

The average ($\overline{a}$) and standard deviation (STD; $\sigma_{a}$) of the parameter $a$ in each $\alpha$ bin of width $\Delta\alpha=0.2$ are shown together in the left panel of Fig.~\ref{fig:std and average of a}.
The former quantifies the average second order shape of SED while the latter characterizes the SED diversity. 
The error bar of $\overline{a}$ is simply $\sigma_{a}$ divided by the square root of star number. 
The most striking feature is that $\sigma_{a}$ nearly monotonically decreases with decreasing $\alpha$, which quantitatively reproduces the same overall trend of diminishing SED diversity revealed in Fig.~6 of \citet{Liu2023}.
This trend uncovers the fact that the SED diversity (perhaps also disk-structure diversity) continues to diminish with disk evolution. 
Furthermore, the nearly linear overall correlation between $\sigma_{a}$ and $\alpha$ hints that the dominant causes of the SED diversity must have some simple relationship with disk thermal structure (traced by $\alpha$) in the entire lifetime of a protoplanetary disk.
Further detailed modeling such as those of \citet{Kimura2016,Schib2021,Karam2023} is needed for a quantitative interpretation.
Distinct from the STD, $\overline{a}$ shows a broad peak in the range of $\alpha<0$ but remains around zero for $\alpha>0$. 
We will interpret these features stage by stage below.
These features indicate that the K-24 SEDs are not dictated purely by random factors, but by the physical relationships of the disk parameters. 
However, another valuable fact is that $\sigma_{a}\ge\overline{a}$ is roughly true for almost all $\alpha$ values. 
Even around the $\alpha\sim-1$ histogram peak, $\sigma_{a}$ is still comparable to the average. 
This means the SED diversity is always significant in all disk evolutionary stages, which cautions us that the average view of disk evolution in terms of K-24 SED may not be simply applicable to individual disks.

Stage B is characterized by very high SED diversity ($\sigma_{a}$) but nearly zero average SED concavity ($\overline{a}$).
This fact simply expresses that there does not exist representative SED in this stage and half of the SEDs are convex and half concave.
Because the YSOs in this early stage very likely still possess a thick envelope with bipolar cavities, inclination angle should be the most sensitive parameter to determine the observed SED shape. 
However, as we mentioned in the beginning of this subsection, if the contamination from edge-on disks is the dominant factor controlling the observed $\alpha$ histogram, 
we should expect a large fraction of YSOs to be edge-on opaque disks that have moved here from Stage C (because of many more stars in Stage C than in B) and they have more concave SED shapes, according to simulations such as \citet[][see e.g., their Fig.~4]{Whitney2003}. 
Thus, in the left panel of Fig~\ref{fig:std and average of a}, the average SED concavity $\overline{a}$ is expected to be larger in this stage than in C, which is not the case in the figure.
Therefore, we can conclude that the member stars in this stage should not be dominated by the contamination from edge-on opaque disks modeled by \citet{Whitney2003}. 
Instead, it can be interpreted by the gradually clearing of the spherical envelope.
We also notice that the $\overline{a}$ curve suffers from large oscillations when $\alpha>1$ but converges better to zero in the range of $1>\alpha>0$. 
This can be understood as the consequence of the increase of star number with decreasing $\alpha$, as can be seen in the left panel of Fig.~\ref{fig: alpha-age}.
This interpretation is also consistent with the larger error bars in those oscillating $\alpha$ bins in the left panel of
 Fig.~\ref{fig:std and average of a}.

Stage C is the right half of the sharp $\alpha\sim-1$ peak. 
While $\sigma_a$ continues to linearly drop with decreasing $\alpha$ in this stage, the average SED concavity $\overline{a}$ first rises up quickly to a peak value of $\sim0.95$ around $\alpha=-0.7$ (i.e., the SEDs become more concave) and then falls down slowly (i.e., become slightly less concave) towards the left border at $\alpha=-1$.
Similarly, the $\alpha>-0.7$ part of this stage cannot be dominated by the contamination from edge-on opaque disks moved here from the $\alpha<-0.7$ part, because otherwise we should see more concave SEDs (with larger $\overline{a}$) in the former part than in the latter, which is not the case in the figure.
The sharp rise of $\overline{a}$ with decreasing $\alpha$ can be interpreted by physical reasons -- very likely the stabilization of the disk to a structure maintaining significant $24\mu$m excessive emission but increasingly less at NIR.
This is possible if a large part of the inner disk is gradually shadowed by the building up of a puffed-up inner rim \citep{Dullemond2001,Dullemond2004a,Dullemond2004} while the outer disk is still flaring (remaining warm by illumination to maintain enhanced $24\mu$m emission).
The establishment of the disk shadow drives the slow decrease of $\alpha$, while the stabilization of disk structure means the slow down of disk evolution, which explains the sharp rise of the $\alpha$ histogram towards the peak at $\alpha\sim-1$.
Then, given that the $\alpha>-0.7$ part of this stage is already not dominated by the contamination from edge-on opaque disks, as discussed above, the $\alpha<-0.7$ part that has many more stars should be even less so.
The slow decrease of the average SED concavity $\overline{a}$ with decreasing $\alpha$ in this part also can be interpreted by physical processes -- possibly the slow photoevaporation of the flaring outer disk and/or gap opening by proto-giant-planets, both of which have the capability to reduce the $24\mu$m emission so as to decrease both the SED slope $\alpha$ and the concavity $a$ to some extent.

Stage D is the left half of the $\alpha\sim-1$ peak. Both the average $\overline{a}$ and dispersion $\sigma_a$ of the SED concavity have stopped the decreasing trends with decreasing $\alpha$.
There is even the indication of a small bump in both curves in the range of $-2<\alpha<-1.4$, which means the K-24 SEDs become slightly more concave and diverse than expected from the trends in Stages B and C.
It is obvious that this stage is not dominated by edge-on disk contamination, given that none of Stages B and C is.
The bump can be interpreted also by physical reasons such as the creation of a large inner hole or wide gaps by giant planets in some disks (the transitional disks) or any yet unknown special properties of the final disk dispersal processes (very likely by photoevaporation). 
In addition, the contamination of the sample by some debris disks is also possible. 

In Stage E, both $\overline{a}$ and $\sigma_{a}$ decrease with decreasing $\alpha$ and $\sigma_{a}>\overline{a}$ is always true.
These trends should be discuss in future when the sample would be more complete in this stage.

We also introduce the disk-SED model grid of \citet{Robitaille2006} to compare with the observed SED concavities, as we did in the comparison with the observed $\alpha$ histogram in Sect.~\ref{subsection:SED slope and age}.
We fit the SED concavity $a$ of all their disk models and show the average and dispersion of the model SED concavities as functions of K-24 SED slope $\alpha$ in the middle panel of Fig.~\ref{fig:std and average of a}. 
It can be seen that the distributions of both the average and dispersion of the model concavities are very different from the observed ones in the middle panel: the model values in both stages B and C are much larger than observations, which means the model-disk parameter distributions can be constrained to some extent by comparing them with the observed ones.
As an illustration of this capability, we show in the right panel of Fig.~\ref{fig:std and average of a} the modeled curves obtained from the simple experiment mentioned in Sect.~\ref{sec:alpha-histo} in which we constrain the models to central star mass $M^*<1$\,M$_\odot$. 
We can see significant changes in both the $\overline{a}$ and $\sigma_{a}$ curves compared to the full grid model in the middle panel.
However, the discrepancies between the constrained models and observations are not only not improved in the stages B and C, but also worsen in stage D.
We envisage that a better strategy involving more disk parameters in a more elaborated manner would yield more useful model constraints, which is however beyond the scope of this paper.

In a summary, the K-24 SED concavity $a$ has allowed us to make a conservative point that, although the inclination angle effect could still be important in all of the Stages B, C and D (because of the always large SED diversity), the contamination from edge-on disks is never the dominating factor in any of these stages.
This suggests that most of those edge-on disks are likely opaque even at IR wavelengths so that most of them actually did not enter the Spitzer/IRAC selected sample.
Both the K-24 SED slope $\alpha$ and concavity $a$ are consistent with the intrinsic evolution of the disk structures in these stages, including the clearing of the spherical envelope in Stage B, the formation and photoevaporation of shadowed flaring disks in Stage C and the development of transitional disks and the final runaway disk dispersal in Stage D.

\section{Test the monochromatic luminosity trends}
\label{sec:luminosity-trends}

\begin{figure*}[!htb]
    \centering
    \includegraphics[width=0.8\textwidth]{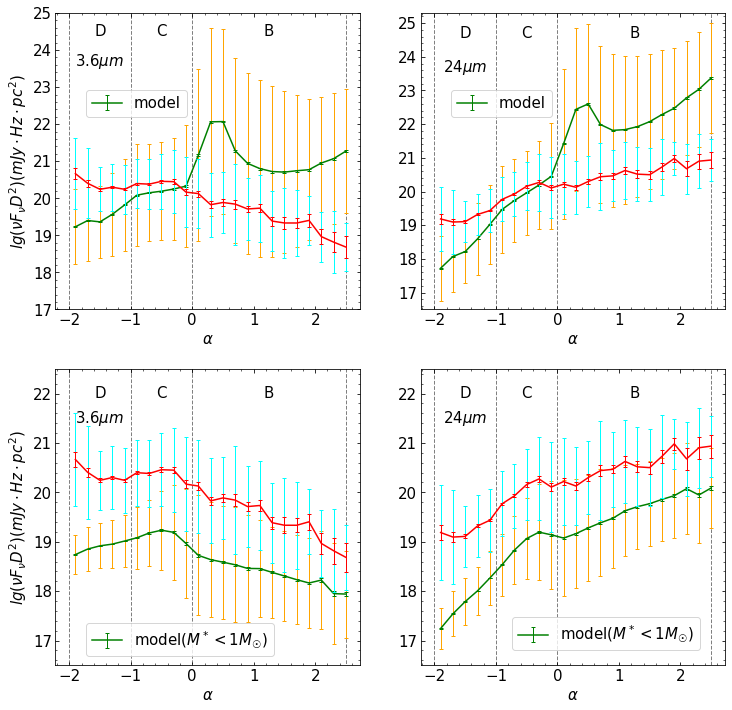} 
    \caption{The distribution of average luminosity $\log_{10}(\nu F_{\nu} D^{2})$ vs K-24 SED slope $\alpha$ (in each $\alpha$ bin of width 0.2) in the 3.6$\mu$m (top left) and 24$\mu$m (top right) bands, respectively,  for the Gould's Belt YSOs (red) and the 200710 disk models of \citet[][green; averaged over the 50 model apertures]{Robitaille2006}, where $D$ is the average Gaia distance to each observed protocluster and 250\,pc for the models. 
    The small error bars show the uncertainty of the average luminosity while the large error bars (cyan for observed YSOs and orange for the models) are the sample dispersion of the luminosities.
    The bottom panels do a similar comparison but with the models confined to those with central star mass $M^*<1$\,M$_\odot$.
    The data of the curves and error bars in the top panels are given on the GitHub mentioned in Abstract.
    }
    \label{fig:flux_alpha}
\end{figure*}

Our next attempt is to examine the statistics of the zeroth order SED feature, the monochromatic IR luminosities of the YSOs. 
We will discuss whether the important factor, the edge-on disk contamination, is a dominating factor and whether the results are consistent with disk structure evolution.
According to the modeling of \citet{Wood2002,Whitney2003}, edge-on opaque disks are not only redder, but much fainter at NIR wavelengths due to the high extinction of the optically thick disk. 
If any sub-region of the $\alpha$ histogram contains a significant fraction of edge-on disks, they may stand out as objects with significantly lower IR luminosity than the others and the average luminosity in that sub-region will be lowered.

For this purpose, we plot the luminosities of the GB YSOs in the IRAC and MIPS\,$24\mu$m bands as functions of the SED slope $\alpha$ in the left column of the Fig.~\ref{fig:luminosity_all} in Appendix~\ref{app:luminosity}.
The luminosity is represented by $\nu F_{\nu} D^{2}$ where $\nu$ is the frequency, $F_\nu$ is the flux and $D$ is the averaged Gaia distance (weighted by signal-to-noise ratio) to each protocluster derived similarly as in \citet{Liu2023}. 
YSOs without reliable $24\mu$m flux are excluded because the version of $\alpha$ used in this work cannot be well defined without it.
To facilitate discussion, we also compute the average of $\log_{10}(\nu F_{\nu} D^{2})$ in each $\alpha$ bin of width 0.2 (red curves) in the Stages B, C and D which we are interested in.
Two sets of error bars are shown, with the larger one (cyan) being the sample dispersion and the smaller one (red) the uncertainties of the averages (dispersion divided by the square root of star number; both defined also in the logarithmic space). 

In order to better understand the statistical behaviors of the observed YSOs, we also repeat the above operations to all the model SEDs of the disk model grid of \citet{Robitaille2006}. 
We compare the average and dispersion of luminosities between the models and observations in the second column of Fig.~\ref{fig:luminosity_all}.

To facilitate discussions here, we only pick out the most interesting $3.6\mu$m and $24\mu$m luminosity curves and show them side by side in the top row of Fig.~\ref{fig:flux_alpha}.
Interesting trends can be seen in the two plots. 
The observed $3.6\mu$m luminosities (red curve) show an overall increasing trend with decreasing $\alpha$.
However, the trend becomes flat in Stages C and D. 
Oppositely, the $24\mu$m luminosities (also red curve) show an overall decreasing trend with decreasing $\alpha$.
The decreasing trend in Stages C and D is slightly steeper than in Stage B. 
Another interesting point is that the larger sets of error bars are almost homogeneous in all $\alpha$ bins despite of the huge difference in star numbers among the bins, which hints that the large scatter in the plots are mainly not due to random measurement errors, but due to intrinsic physical diversities of the YSOs in all three stages.
Possible contributors to the luminosity diversity include different stellar and disk masses and intermittent stellar accretion.
Our analysis of the apparent fluxes (not plotted here) shows that the detected YSOs all have their IRAC and MIPS\,$24\mu$m fluxes far above the nominal sensitivity limits of the Spitzer bands given by \citet{Werner2004}, which means that the luminosities in Fig.~\ref{fig:flux_alpha} do not suffer from sensitivity limitation.

These results do not support the contamination from edge-on disks to be important in the sample.
At the level of individual stars, we do not see the expected subsample of YSOs standing out as distinctive low luminosity sources in any of the three stages.
Although we cannot exclude the existence of some edge-on opaque disks, they must not be numerous in the observed sample.
At the population level, if Stage B is strongly contaminated by edge-on opaque disks moved there from Stage C where there are many more member stars than in B, due to the strong disk extinction, we should expect the Stage B YSOs to have significantly lower luminosities than those in C. 
However, at least the $24\mu$m luminosities in the right panel of the figure show an opposite trend which should be interpreted by stronger excessive $24\mu$m emission from the more massive cool spherical envelope in Stage B. 
The grand anti-correlation between the $3.6\mu$m luminosity and $\alpha$ in Stage B can be explained by either the dominance of edge-on opaque disk contamination or the evolution of the YSOs (i.e., progressively clearing the envelope) or both, because both factors function through the change of opacity (reddening).
However, only the latter is consistent with the trend in the $24\mu$m luminosities.
Similarly, were Stage C dominated by edge-on disks moved there from Stage D due to strong extinction of edge-on disks, the YSOs in this stage should have significantly lower luminosities than those in Stage D, but this is not the case in either panel of the figure.

After excluding the contamination from edge-on disks, we shall examine whether the luminosity trends in Fig.~\ref{fig:flux_alpha} are consistent with disk evolutionary processes.
Both the $3.6\mu$m and $24\mu$m luminosity trends in Stage B have been reasonably suggested in the previous paragraphs to be driven by the clearing of the envelope. 
Further on, the comparable $3.6\mu$m luminosities in Stages C and D can be due to the slow evolution of the stellar luminosity, because stellar emission has become the major contributor in these stages.
The more steeply decreasing $24\mu$m  luminosities with decreasing $\alpha$ in these two stages can be explained by the formation of giant planets that open broad gaps to destroy the part of disks that is mainly responsible for the $24\mu$m emission or by the photoevaporation of the outer disks.
It is worth noting that the transition point at $\alpha=-0.3$ in both panels of the figure is close to the $\alpha$ value where the average K-24 SED concavity $\overline{a}$ rises up quickly from 0 to $\sim1$ as the $\alpha$ decreases in Fig.~\ref{fig:std and average of a}, which implies that they may be two manifestations of the same process: the start of giant planet formation.

The luminosity curves of the disk model grid of \citet{Robitaille2006} in the top row of Fig.~\ref{fig:flux_alpha} (green curves) show increasing trends with increasing $\alpha$ value and a peak around $\alpha=0.4$ in both 3.6 and 24$\mu$m bands.
These trends are in opposite sense than the observed 3.6\,$\mu$m luminosity distribution and significantly steeper than the observed 24\,$\mu$m luminosity trend.
In addition, the sample dispersion of the model grid (error bars in orange) are definitely larger in the $\alpha>0$ range than in the $\alpha<0$ range, which is also distinct from the uniform sample dispersion (error bars in cyan) of the observed luminosities. 
These large differences indicate that the luminosities can provide meaningful constraints to disk model grids as well. 
To illustrate this potential, we show in the bottom row of Fig.~\ref{fig:flux_alpha} the distributions of the modeled 3.6 and 24\,$\mu$m luminosities (green curves) in the aforementioned simple experiment in which we constrain the model grid to central star mass of $M^*<1$\,M$_\odot$.
Interestingly, the distributions of the low mass YSO models show much more similar trends as the observed ones (red curves). 
However, the modeled luminosities are collectively lower than the observed ones by about an order of magnitude.
It is clear that there is the room to further constrain the disk parameters to improve the observation-model agreement, which is beyond the scope of the current work.

In a summary, the monochromatic IR luminosities do not support the contamination from edge-on opaque disks to be a dominant factor in Stages B, C and D in the $\alpha$ histogram. {Instead, they are consistent with disk evolution.} 
This conclusion casts some doubt on the high fraction of more than $30\%$ of mis-classified Class\,\textsc{ii} edge-on disks in the model of \citet{Crapsi2008}. 
Possible improvements to their model are to consider round inner disk wall as suggested by \citet{Isella2005}, to consider optically thicker disks so that edge-on disks may become too faint to enter the observed sample, or alternatively to consider more transparent disks so that many edge-on disks have little impact to K-24 SEDs, or to assume a geometrically thinner disk.
On the other hand, the nearly uniform, large, luminosity scatter in all stages in Fig.~\ref{fig:flux_alpha} also warn us that the $\alpha$ histogram is mainly suitable for the `average disk' while applying it to individual stars must take into account the large intrinsic diversity.

\section{Large uncertainty in the SED-slope histogram peak?} 
\label{sec:Remaining SED randomness}

The $\alpha$ histogram peak around $-1$ mimics a Gaussian function to which one cannot help suspecting that it might be still dominated by random factors, instead of by disk evolution.
Particularly, the large intrinsic diversity of the disks revealed in Figs.~\ref{fig:std and average of a} and \ref{fig:flux_alpha} hints that the sample dispersion may not be thoroughly compressible through averaging.
However, if we assume that the peak shape is still dominated by randomness, the underlying true $\alpha$ histogram of the average disk would be even narrower than currently observed. 
This hints at a more extreme disk evolution pattern in which the K-24 SED slope $\alpha$ of a given YSO must evolve from a larger value (e.g., 0) into the peak value of $\sim-1$ in a sudden jump, which is not likely for huge objects like the protoplanetary disks.
Furthermore, the slight shift of the histogram peak from $\alpha=-1$ to the asymmetric position $\alpha=-1.03$ in the improved version of the $\alpha$ histogram in Fig.~\ref{fig: alpha-age} also hints that this peak is not a perfect Gaussian function.

\section{Summary} \label{sec:summary}

We have critically examined the uncertain factors behind the histogram of K-24 SED slopes $\alpha$ of Gould's Belt YSOs. 
We have to confess that still little is known of the statistical nature of most of the complicated physical factors involved in the evolution of protoplanetary disks so that they have to be assumed to be random factors for the moment.
It is obvious that the $\alpha$ histogram cannot trace disk evolution well in the earliest Stage A and the latest Stage E due to sample incompleteness. 
By using additional information from monochromatic luminosities and SED concavity, we are able to show that, although the disk inclination angle might be always important, the contamination from edge-on opaque disks is never a dominating factor to the observed K-24 SED at a population level. 
The statistics of the luminosity, SED slope $\alpha$ and concavity can trace the overall evolutionary sequence of an `average disk' that includes envelope clearing, formation of self-shadowed flaring disk, gap or hole opening by planet formation, photoevaporation etc. 
A linear decrease of K-24 SED diversity with decreasing $\alpha$ is found (left panel of Fig.~\ref{fig:std and average of a}), but needs further modeling to explain quantitatively.
However, the application of these results to individual stars should be done only with an adequate account for the large intrinsic diversities of the IR tracers that are not necessarily purely random in nature.

\normalem
\begin{acknowledgements}
ML jointly motivated this work, analyzed the data and drafted the manuscript; JH jointly motivated this work, participated in the data analysis and revised the manuscript; the other co-authors discussed and improved many parts of the work and manuscript.
This work is supported by the Natural Science Foundation of Yunnan Province (No. 202201BC070003).
ZG is supported by the ANID FONDECYT Postdoctoral program No. 3220029. ZG acknowledges support by ANID, -- Millennium Science Initiative Program -- NCN19\_171.
This work is sponsored (in part) by the Chinese Academy of Sciences (CAS), through a grant to the CAS South America Center for Astronomy (CASSACA) in Santiago, Chile.
\end{acknowledgements}

\appendix

\section{Improvements to the histogram of SED slopes} \label{subsec:alpha_histograms}

\subsection{The improvements} 
\label{subsec:ImproveHisto}

We have carefully examined the procedures of \citet{Liu2023} and found that the following tow aspects deserve improvements: 1) the resolution of the average $\alpha$ histogram is far from optimal; 2) their choice of averaging the $\alpha$ histograms among the 13 protoclusters is only one of the possible schemes to derive the average $\alpha$ histogram, while the impact of different schemes to the resulting average $\alpha$ histogram is still unclear.

\begin{figure}
    \centering
    \includegraphics[width=1\textwidth,trim=0cm 0 0 0,clip]
    {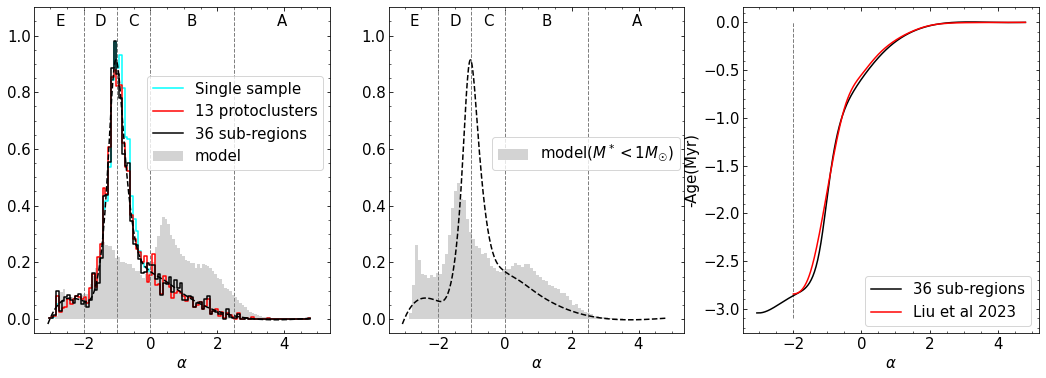} 
    \caption{
    The left panel shows the $\alpha$ histograms derived with three star grouping schemes: 
    1) taking all the 5194 GB YSOs as a single sample (cyan); 2） averaging the histograms over the 13 largest GB protoclusters (red; 4909 YSOs in total); 
    3) averaging the histograms over the 36 sub-clusters (black; 4312 YSOs in total). 
    The black dashed curve is a smoothed cubic spline fitting to the best histogram, the 36-sub-cluster case. 
    The gray shaded histogram is the $\alpha$ histogram of the 200710 disk models of \citet[][SED averaged over the 50 model apertures]{Robitaille2006}.
    The vertical dashed lines show the five disk evolution stages suggested by \citet{Liu2023}.
    The middle panel compares the smoothed histogram in the left panel with the model histogram delimited only to those models with central star mass $M^*<1$\,M$_\odot$.
    The right panel shows the new $\alpha$-to-disk age conversion curve (black) derived from the smoothed $\alpha$ histogram in the left panel, which can be compared to the result from \citet[][red curve]{Liu2023}. The vertical dashed line in the right panel marks the critical value of $\alpha=-2$ to the left of which our sample becomes incomplete. The data of the smoothed $\alpha$ histogram and the new alpha-to-age conversion curve are given on the GitHub mentioned in Abstract.}
    \label{fig: alpha-age}
\end{figure}

There are several empirical approaches to determine the optimal histogram bin width of specific kinds of sample distributions. 
Usually the optimal bin width can be smaller for a larger sample size. 
If the random samples are averaged in sub-groups before final averaging, the effective sample size could be slightly smaller than the total sample size, depending on the weighting among sub-groups. 
However, this shrinkage of sample size is not significant and can be neglected, according to our practice on the sample in this work.
Therefore, the choice of the `Freedman-Diaconis' rule \citep{Freedman1981OnTH} which is good enough for a singly peaked histogram can be justified for the work of \citet{Liu2023}.
However, they arbitrarily adopted a conservative bin width of $\Delta\alpha=0.5$ without any explanation. 
If we adopt the total size of 5194 YSOs in all the 28 GB protoclusters surveyed by Spitzer Space Telescope, we derive the optimal bin width of $\sim0.082$.
As can be seen from the left panel of Fig.~\ref{fig: alpha-age}, the higher $\alpha$ resolution still produces a smooth histogram but presents more details of the distribution.

There are also different ways to make the $\alpha$ histogram for a large sample of YSOs. 
The most straightforward way is to merge all the YSOs of the 28 GB protoclusters surveyed by Spitzer Space Telescope into a single sample. 
A drawback of this approach is that large protoclusters, such as Orion A, may dominate the resulting histogram so that it may be biased by the special properties  (e.g., the temporal fluctuation of star formation rate) of few largest clusters.
\citet{Liu2023} adopted the natural star grouping of the individual GB protoclusters and averaged the histograms over the 13 largest ones to improve the imbalance of cluster weights. 
However, the 13 clusters still have a broad range of sizes from 70 to 2435 member stars.
Some of the large protoclusters can be broken into smaller sub-clusters.
Therefore, we visually assess the clustering of member stars in each of the 13 GB protoclusters in the sky plane and also in their distribution of $\alpha$ value and further divide them into sub-clusters (36 in total; the details will be given in a future paper). 
These sub-clusters still show similar $\alpha$ histograms, but with the differences among them being larger than among the 13 protoclusters.

We compare the resulting (averaged) $\alpha$ histograms from the three approaches in the left panel of Fig.~\ref{fig: alpha-age} to assess the impact. 
The histogram made by taking all the GB YSOs as a single sample (cyan) shows a peak position at a slightly larger $\alpha$ value than the other two, very likely because it is dominated/biased by the largest cluster, Orion A, which has contributed nearly one half of member stars in the `single sample' case.
The histograms from the 13 largest GB protoclusters (red) and from the 36 sub-clusters (black) are more similar in overall shape, which hints that more deliberately grouping the YSOs may not benefit further.  
However, there still are small but potentially important differences between the latter two: 
1) the 36-sub-clusters histogram shows a stronger and narrower peak around $\alpha=-1.03$ which best informs an intriguing asymmetry in the peak shape and
2) shows a slightly stronger secondary peak around $\alpha=-2.5$.
Because the grouping of the 36 sub-clusters has given the diverse sub-clusters the most homogeneous weights in the averaging, we consider it as the best approach.

Similar as \citet{Liu2023}, we fit a cubic spline function (black dashed, left and middle panels of Fig.~\ref{fig: alpha-age}) to the best $\alpha$ histogram using the function \textit{splrep} of the Python package \textit{scipy.interpolate}, with an empirical choice of the smallest possible smoothing parameter $s=0.08$ to suppress unrealistic fluctuations in the fitted curve.
Then, we integrate the fitted $\alpha$ histogram to derive an $\alpha$-to-age conversion curve, as shown in the right panel of Fig.~\ref{fig: alpha-age} (black curve).
Compared to the version of \citet[][red curve]{Liu2023}, our new curve is steeper around the $\alpha=-1$ peak and extends down to the incomplete range of $\alpha=-3\sim-2$.

\subsection{The new SED-slope histogram}
\label{subsection:SED slope and age}

Now we discuss the new features of the updated $\alpha$ histogram and re-examine its potential to trace disk age. 
First of all, if we adopt the same criteria of \citet{Liu2023} to divide evolutionary stages according to the characteristics of the $\alpha$ histogram, the five evolutionary stages, A, B, C, D and E as proposed by them and shown in the left and middle panels of Fig.~\ref{fig: alpha-age} still roughly agree with the statistical features of the new $\alpha$ histogram. 
Thus, we do not tend to update it.

The Stage A ($\alpha>2.5$) involves very few observed YSOs. 
The histogram in this stage is not reliable and thus cannot trace the evolutionary age well. 
Fortunately, however, this stage is likely very brief in nature so that its impact to the estimation of disk age in more evolved stages is limited.

In Stage B ($2.5\ge\alpha>0$), our new $\alpha$ histogram shows a clearer ramp-up with decreasing $\alpha$ than in that of \citet{Liu2023}.
The nearly linear shallow distribution may allow the $\alpha$ histogram to trace disk evolution in a simple linear manner. 
However, the total number of YSOs involved in this stage is still small so that the impact of large uncertainties still need to be further discussed (see in next section). 

Stages C ($0\ge\alpha>-1$) and D ($-1\ge\alpha>-2$) bracket the major peak of the $\alpha$ histogram. 
However, with the finer resolution, this peak is definitely sharper and taller than in that of \citet{Liu2023}. 
In addition, we also notice a small shift of the peak position to $-1.03$, which indicates some asymmetry in the shape of the peak. 
If confirmed by future observations, the asymmetry would indicate that the clearing of the primordial disk is faster than establishing it.

The peak of the $\alpha$ histogram around -1 has an interesting indication. 
It is known that a perfectly flat disk heated by UV radiation from an extended UV emission region around the central star or dominated by steady disk-accretion heating has a theoretical K-24 SED slope of $-1.33$ \citep{Dullemond2007}. 
The observed $\alpha$ peak of $-1$ (or more exactly $-1.03$) is significantly larger, which means the disk structure in the most stable evolutionary phases must have non-negligible flaring degree.
This conclusion also agrees to later observations \citep[e.g.,][]{Avenhaus2018}.
\citet{Dullemond2007} showed that a disk model with an K-24 SED slope $\alpha=-1$ should assume a radial temperature profile of $T_{eff} \propto r^{-\frac{2}{3}}$, which is shallower that of a flat disk model, $T_{eff} \propto r^{-\frac{3}{4}}$.

Concerning $\alpha$ as a disk evolution tracer, even if this is true, the sharp peak of its histogram indicates that the disk must evolve very slowly near the peak so that the $\alpha$ values within the full width at half maximum (FWHM) of the peak ($-1.3 < \alpha < -0.7$) might not be a sensitive tracer of the disk evolution. 
Nevertheless, as we will show in Sect.~\ref{sec:luminosity-trends}, even within the FWHM of the peak, $\alpha$ still has the potential to trace the disk evolution to some extent.

The completeness of samples in Stage E is not great because only few protoclusters have their older Class III member stars \citep{Liu2023} successfully identified, so it is difficult for $\alpha$ to accurately reflect the age of disk evolution.
Due to the same reason, we also do not discuss the secondary peak around $\alpha=-2.5$.

To check whether the observed $\alpha$ histogram can be used to constrain model grids in a statistical sense, we compare it to the disk model grid of \citet{Robitaille2006}.
We compute the K-24 SED slope $\alpha$ for all the 200710 disk models in the same manner as for the observed SEDs and show their histogram in the left panel of Fig.~\ref{fig: alpha-age} (gray filled).
The model histogram shows two broad peaks, which is distinct from the observed single sharp peak around $\alpha=-1$. This striking difference demonstrates the usefulness of the observed $\alpha$ histogram to constrain future disk evolution model grids.
As an illustration of this capability, we perform a simple experiment by constraining the models to those with central star mass $M^*<1$\,M$_\odot$, because it is well known that the GB YSOs are dominantly low mass stars.
The constrained model $\alpha$ histogram is shown in the middle panel of Fig.~\ref{fig: alpha-age} (gray filled).
Compared with the one of the full model grid in the left panel (gray filled), the peak in the $\alpha>0$ region is now greatly dwarfed, reflecting the disappearance of massive-central-star models and indeed better agreeing to the observed $\alpha$ histogram (black dash curve).
The remaining model $\alpha$ histogram peak is sharper, but still peaked around $\alpha\sim-1.5$ which is significantly smaller than the observed peak value.
It is encouraging to further refine the model parameters to achieve a better agreement, which is however beyond the scope of this paper.

In a summary, solely judged from the properties of the derived average $\alpha$ histogram itself, the K-24 SED slope $\alpha$ may not be a reliable tracer of disk evolution in the earliest Stage A, the earlier part of Stage B, the last Stage E, and even in the core part of the $\alpha=-1.03$ peak. 
However, Stages B, C and D of the histogram still can provide useful constraints to future disk evolution model grids.

\section{All luminosity plots}\label{app:luminosity}

\begin{figure*}[!htb]
    \centering
    \includegraphics[width=0.49\textwidth]{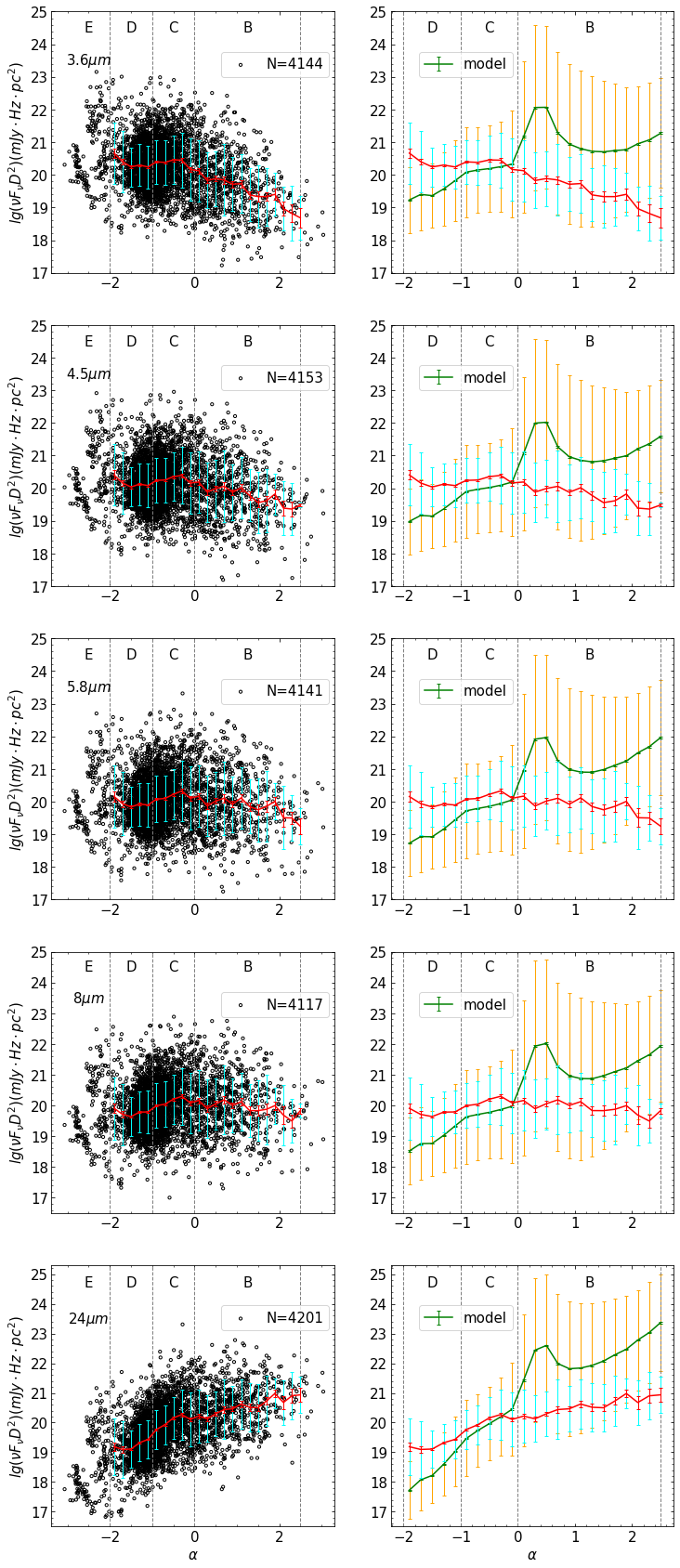} \includegraphics[width=0.49\textwidth]{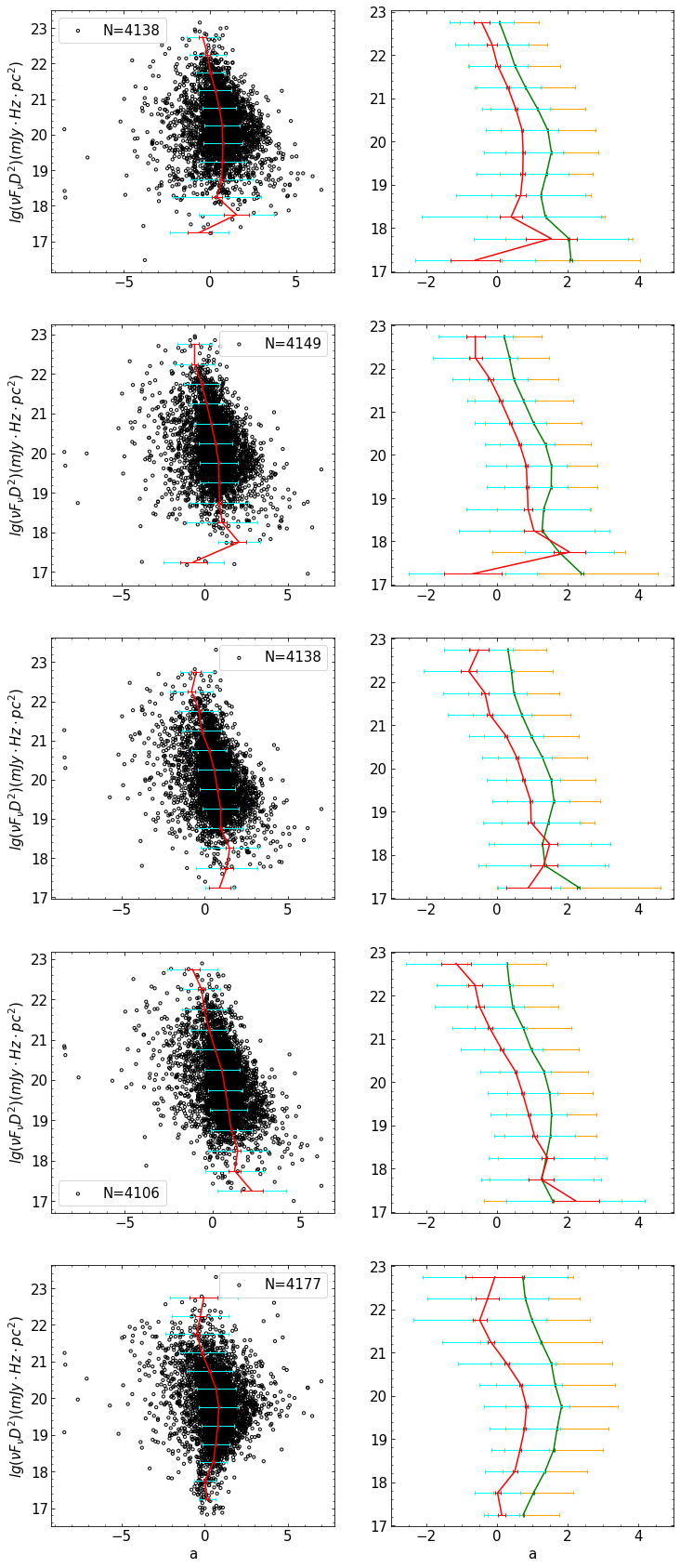} 
    \caption{The distribution of observed luminosity $\log_{10}(\nu F_{\nu} D^{2})$ in the Spitzer IRAC and MIPS\,$24\mu$m bands vs K-24 SED slope $\alpha$ and concavity $a$ for all the YSOs in Gould's Belt (first and third columns) and the comparison with the similar distributions of the 200710 disk models of \citet{Robitaille2006} (second and fourth columns; averaged over the 50 model apertures), where $D$ is the average Gaia distance to each observed protocluster and $D=250$\,pc for the models. 
    The red curve and error bars show the average observed quantities (luminosity or concavity) and its uncertainty in each $\alpha$ or luminosity bin, while the cyan error bars show the sample dispersion in each bin. The meanings are similar for the model curves (green) and error bars (green and orange; the green error bars are usually very small) in the second and fourth column. 
    The data of the curves and error bars are given on the GitHub mentioned in Abstract.
    }
    \label{fig:luminosity_all}
\end{figure*}

The correlations between all the observed monochromatic luminosities in the four IRAC and the MIPS\,24$\mu$m bands with the corresponding K-24 SED slope $\alpha$ are presented in the left most column of Fig.~\ref{fig:luminosity_all}. 
A comparison with similarly computed monochromatic luminosities of the disk-SED model grids of \citet{Robitaille2006} is given in the second column of the figure.
See the detailed discussions of them in Sect.~\ref{sec:luminosity-trends}.
However, we omit the 2MASS Ks band here because the ground based photometry is sensitivity limited, while the Spitzer observations are free of this impact in all the five used bands. 
This issue should be investigated in more detail in future works when 2MASS photometry would be needed to constrain disk evolution models.
In addtion, for completeness, we also plot similar correlations between the luminosities and the SED concavity $a$ among the observed YSOs and among the models of \citet{Robitaille2006} in the third and fourth columns of the Fig.~\ref{fig:luminosity_all}. 
They may be useful for constraining future model grids. 
We do not discuss the details here.

\bibliographystyle{raa}
\bibliography{ms2024-0114}




\end{CJK*}

\end{document}